# A digital score of tumour-associated stroma infiltrating lymphocytes predicts survival in head and neck squamous cell carcinoma

Muhammad Shaban[1], Shan E Ahmed Raza[1], Mariam Hassan[2], Arif Jamshed[2], Sajid Mushtaq[2], Asif Loya[2], Nikolaos Batis[3], Jill Brooks[3], Paul Nankivell[3], Neil Sharma[3], Max Robinson[4], Hisham Mehanna[3], Syed Ali Khurram[5], Nasir Rajpoot[1,6,7*]

[1] Department of Computer Science, University of Warwick, UK

[2] Shaukat Khanum Memorial Cancer Hospital Research Centre, Lahore, Pakistan

[3] Institute of Head and Neck Studies and Education, University of Birmingham, UK

[4] School of Dental Sciences, Faculty of Medical Sciences, Newcastle University, Newcastle upon Tyne, UK.

[5] School of Clinical Dentistry, University of Sheffield, Sheffield, UK

[6] The Alan Turing Institute, London, UK

[7] Department of Pathology, University Hospitals Coventry & Warwickshire NHS Trust, Coventry, UK

* n.m.rajpoot@warwick.ac.uk

**Abbreviations:** TILs, Tumour Infiltrating Lymphocytes; TAS, Tumour Associated Stroma; TASIL, Tumour-Associated Stroma Infiltrating Lymphocyte; DSS, Disease-specific Survival; DFS, Disease-free Survival

TASIL-score predicts survival in head and neck squamous cell carcinoma

## Abstract

The infiltration of T-lymphocytes in the stroma and tumour is an indication of an effective immune response against the tumour, resulting in better survival. In this study, our aim is to explore the prognostic significance of tumour-associated stroma infiltrating lymphocytes (TASILs) in head and neck squamous cell carcinoma (HNSCC) through an AI based automated method. A deep learning based automated method was employed to segment tumour, stroma and lymphocytes in digitally scanned whole slide images of HNSCC tissue slides. The spatial patterns of lymphocytes and tumour-associated stroma were digitally quantified to compute the TASIL-score. Finally, prognostic significance of the TASIL-score for disease-specific and disease-free survival was investigated with the Cox proportional hazard analysis. Three different cohorts of Haematoxylin & Eosin (H&E) stained tissue slides of HNSCC cases (n=537 in total) were studied, including publicly available TCGA head and neck cancer cases. The TASIL-score carries prognostic significance ($p$=0.002) for disease-specific survival of HNSCC patients. The TASIL-score also shows a better separation between low- and high-risk patients as compared to the manual TIL scoring by pathologists for both disease-specific and disease-free survival. A positive correlation of TASIL-score with molecular estimates of CD8+ T cells was also found, which is in line with existing findings. To the best of our knowledge, this is the first study to automate the quantification of TASIL from routine H&E slides of head and neck cancer. Our TASIL-score based findings are aligned with the clinical knowledge with the added advantages of objectivity, reproducibility and strong prognostic value. Although we validated our method on three different cohorts (n=537 cases in total), a comprehensive evaluation on large multicentric cohorts is required before the proposed digital score can be adopted in clinical practice.

## Introduction

There are around 900,000 annual new cases of head and neck cancers worldwide and 450,000 annual deaths [1]. Head and neck squamous cell carcinoma (HNSCC) accounts for approximately 90% of head and neck cancers and is the sixth leading cancer by incidence worldwide [2]. HNSCC predominantly develops in the epithelial lining of the oral cavity, sinonasal tract, pharynx, and larynx [3]. Major risk factors include tobacco smoking, tobacco chewing, alcohol consumption and human papillomavirus infection [4], [5]. The prognosis of HNSCC remains poor with a 28-67% chance of survival at five years [6] highlighting the need for novel biomarkers and objective quantitative analysis of any potential prognostic markers to stratify patients into appropriate risk groups and identify those who may benefit from aggressive treatment from those that can be put under surveillance [7].

Tumour infiltrating lymphocytes (TILs) have been shown to be of prognostic significance for HNSCC [8], [9, p. 3]. TILs are not routinely quantified in diagnostic practice although some methods for manual TIL scoring on Haematoxylin and Eosin (H&E) stained tissue sections have been reported in the literature. However, this quantification process is subjective and prone to inter- and intra-observer variability. Recently, the International Immuno-Oncology Biomarker Working Group has developed guidelines for TIL assessment in breast cancer [10] to standardise and obtain a more reproducible and objective TIL score. However, there are no such guidelines for TIL assessment in HNSCC. Therefore, a computer-based automated method may help to eliminate the subjectivity through objective TIL quantification and assist with prognostic stratification.

The emerging area of computational pathology has seen a surge of interest in recent years with a variety of algorithms proposed in the literature for detection of lymph node metastasis in breast cancer [11], tumour detection [12] and segmentation [13]–[15], cancer grading [16], [17] and prognostics [18], [19], to list only a few. There have been some studies on automated quantification of lymphocytic infiltration in whole slide images (WSIs) of H&E stained histology tissues slides of different cancers. For instance, Saltz *et al.* [20] quantified the spatial patterns of lymphocytes in WSIs, independent of tumour location, to investigate their prognostic significance in fourteen different cancer types. Maley *et al.* [21] reported immune-cell-cancer colocalisation as a prognostic factor for breast cancer. Nawaz *et al.* [22] used hotspot analysis to identify the statistically significant regions of cancer and immune cells. Shaban *et al.* [19] quantified the abundance of TILs for disease-free survival (DFS) analysis of oral squamous cell carcinoma (OSCC) patients. Most existing automated quantification methods either only consider lymphocytes or lymphocytic infiltration in tumour regions. However, some clinical studies [23], [24] have reported the prognostic significance of lymphocyte infiltration in tumour-associated stroma (TAS) in HNSCC.

We propose a novel deep learning based objective quantification method of lymphocytic infiltration in TAS (see Figure 1). The proposed method calculates the percentage of TAS colocalised with lymphocytes that we term as the TASIL-score. We evaluate the prognostic significance on three independent patient cohorts (n=537 cases in total). The proposed TASIL-score shows prognostic significance ($p$=0.002) for disease-specific survival (DSS) of HNSCC patients. The TASIL-score is also a prognostic indicator for DSS and DFS in two OSCC and oropharyngeal squamous cell carcinoma (OPSCC) cohorts. We have also compared the predictive ability of TASIL-score based survival model with existing quantification methods through a concordance index measure where the TASIL-score achieved the highest concordance compared to its counterparts. The main highlights of this work are as follows:

- A novel digital score of tumour-associated stroma infiltrating lymphocytes (TASIL-score) carries prognostic significance for DSS of oral, oropharyngeal, and head and neck SCC patients.
- The TASIL-score is a strong prognostic indicator for DSS and DFS of OSCC and OPSCC.
- The TASIL-score shows a better separation between low- and high-risk patients compared to manual TIL scoring by pathologists for DSS and DFS.
- The TASIL-score also shows a positive correlation with molecular estimates of CD8+ T cells, which is in agreement with existing findings [23].

TASIL-score predicts survival in head and neck squamous cell carcinoma

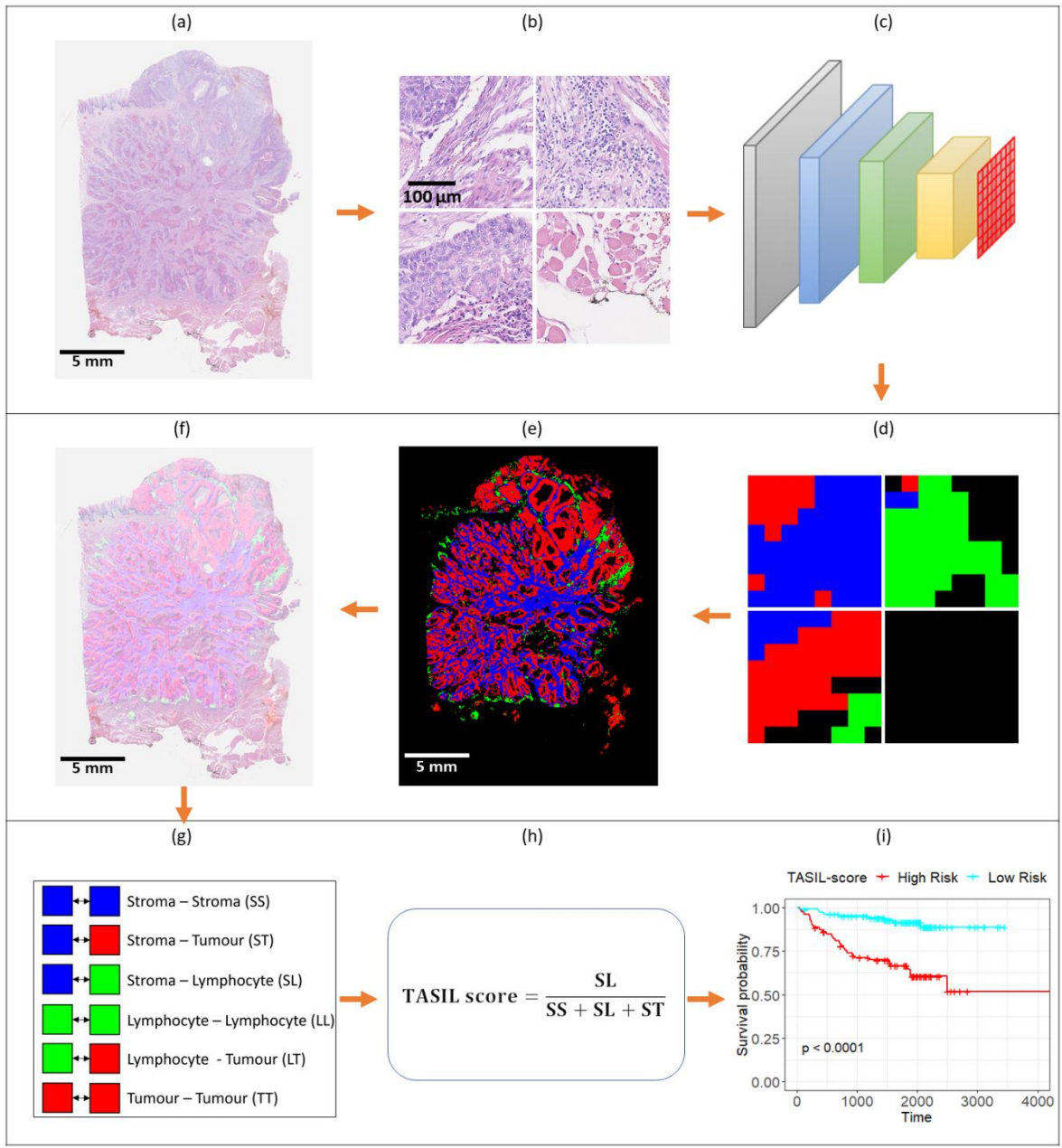

*Figure 1: Flow diagram of the proposed method. (a) A whole slide image (WSI). (b) Image tiles/patches extracted from the WSI. (c) a convolutional neural network based patch segmentation method. (d) Patch based segmentation results where red, green, blue, and black colours represent tumour, lymphocyte, tumour associated stroma, and non-regions of interests, respectively. (e) Segmentation map of the WSI. (f) Overlay of segmentation map over WSI. (g) Analysis of the spatial co-occurrence of tumour, lymphocyte, and tumour-associated stroma regions. (h) Calculation of the proposed TASIL-score based on the number of each pair of co-occurrences. (i) TASIL-score based patient stratification into low and risk groups.*

## Materials & Methods

### Clinical Samples

Three different patient cohorts (TCGA-HN [25], SKM [19] and PredicTR1 [26]) are used in this study. The TCGA-HN cohort (C1) is a publicly available dataset which comprises 450 diagnostic H&E WSI of squamous cell carcinoma (SCC) cases from different sites of the head and neck (H&N). However, many of these slides suffer from preparation and scanning artefacts. After excluding cases with poor quality of slides, our final TCGA-HN cohort consists of 342 cases with one WSI per case. The H&E WSIs from these cases have mostly been scanned at 40× with some scanned at 20×. The SKM cohort (C2) comprises of 100 OSCC cases collected from the Shaukat Khanum Memorial Hospital and Research Centre (SKMCH&RC) Lahore, Pakistan [19]. The PredicTR1 cohort (C3) contains 95 OPSCC cases collected from six different centres across the United Kingdom [26]. The representative H&E tissue sections for both SKM and PredicTR1 cohort were all scanned at 40× magnification.

### Survival Information

The DSS information was available and obtained for most of the selected cases in all three cohorts. The DSS time is calculated from the date of diagnosis to the date of death or the date of the last follow-up in case of censored data. The DFS information was only available for C2 and C3 cohorts. The DFS is censored at the date of first recurrence or death, whichever occurred first, or the date of the last contact for the patients alive without recurrent disease. A detailed description of patient characteristics and available clinical and pathological parameters can be found in the supplementary materials document.

### Quantification of tumour-associated stroma infiltrating lymphocytes

We propose an objective and automated score for the quantification of tumour-associated stroma infiltrating lymphocytes, namely the TASIL-score, which quantifies lymphocytes in the vicinity of TAS using the spatial co-occurrence statistics of both TAS and lymphocytes in a WSI. The TASIL-score is

TASIL-score predicts survival in head and neck squamous cell carcinoma computed in two steps: segmentation of WSIs into clinically significant tissue types and calculation of the spatial co-occurrence statistics.

In the first step, we divided a given WSI into small sub-images (tiles or patches) and employed a deep learning based patch classification method to segment the WSI into four different types of regions: tumour, TAS, lymphocytes, and all other tissue regions as non-regions of interest (Non-ROIs). In the second step, spatial co-occurrence statistics are calculated using the co-occurrence analysis of different types of patches in a WSI. A patch adjacent to another patch in any direction was considered as an instance of co-occurrence. First, six different patch co-occurrence patterns are defined based on three clinically significant tissue types, as shown in Figure S1. Then the TASIL-score is calculated as follows:

$$\text{TASIL-score} = \frac{SL}{SS + SL + ST} \quad (1)$$

where $SL$ represents the number of times tumour associated stroma and lymphocytes patches co-occur in a WSI. Similarly, $SS$ and $ST$ denote the number of co-occurrences of TAS patches with other TAS patches and tumour patches, respectively. The TASIL-score ranges from 0 to 1, where 0 represents no infiltration and 1 represents high degree of lymphocytic infiltration in the tumour associated stroma.

Statistical Methods and Data Analysis

Survival analysis was performed with DSS and DFS data. The Kaplan–Meier estimator was used and the log-rank test performed to stratify patients into low- and high-risk groups (log-rank test based *p*-values were calculated to assess the significance of the various features including the proposed TASIL-score). The Cox proportional hazard regression model was used for univariate and multivariate analyses and 95% confidence intervals were computed. Spearman rank-order correlation coefficient was used for correlation analyses between TASIL-score and molecular estimates, and between TASIL-computed in two steps: segmentation of WSIs into clinically significant tissue types and calculation of the spatial co-occurrence statistics.

In the first step, we divided a given WSI into small sub-images (tiles or patches) and employed a deep learning based patch classification method to segment the WSI into four different types of regions: tumour, TAS, lymphocytes, and all other tissue regions as non-regions of interest (Non-ROIs). In the second step, spatial co-occurrence statistics are calculated using the co-occurrence analysis of different types of patches in a WSI. A patch adjacent to another patch in any direction was considered as an instance of co-occurrence. First, six different patch co-occurrence patterns are defined based on three clinically significant tissue types, as shown in Figure S1. Then the TASIL-score is calculated as follows:

$$\text{TASIL-score} = \frac{SL}{SS + SL + ST} \quad (1)$$

where $SL$ represents the number of times tumour associated stroma and lymphocytes patches co-occur in a WSI. Similarly, $SS$ and $ST$ denote the number of co-occurrences of TAS patches with other TAS patches and tumour patches, respectively. The TASIL-score ranges from 0 to 1, where 0 represents no infiltration and 1 represents high degree of lymphocytic infiltration in the tumour associated stroma.

Statistical Methods and Data Analysis

Survival analysis was performed with DSS and DFS data. The Kaplan–Meier estimator was used and the log-rank test performed to stratify patients into low- and high-risk groups (log-rank test based *p*-values were calculated to assess the significance of the various features including the proposed TASIL-score). The Cox proportional hazard regression model was used for univariate and multivariate analyses and 95% confidence intervals were computed. Spearman rank-order correlation coefficient was used for correlation analyses between TASIL-score and molecular estimates, and between TASIL-score and the manual TIL score assigned by a pathologist. The concordance statistics were used to compare the different automated quantification scores for DSS analysis.

# Results

### Automated segmentation of Whole Slide Images for quantification of TASIL-score

A deep learning based segmentation algorithm was trained and evaluated on WSIs from C1 and C2 cohorts, where 10 WSIs were used for training and 2 for validation from each cohort. More than 179K patches (141K for training and 38K for validation) were annotated by an expert pathologist (SAK). The segmentation method achieved an average accuracy of 0.85 and macro F1-score of 0.83. Quantitative and visual results for WSI segmentation are presented in Table S5 and Figure S2 in the Supplementary Materials. Performance of the segmentation method was further evaluated by calculating the Spearman correlation between the percentage of predicted lymphocyte patches (L-Percentage) and the pathologists' manually assigned TIL score on the C3 cohort. The Spearman correlation score of 0.71 with a highly significant *p*-value of $5.10 \times 10^{-16}$ was observed between the two scores. The distribution of L-Percentage across the three groups of the pathologists' TIL score shows clear separation (see Figure S3).

### Higher TASIL-score is associated with better disease-specific survival of HNSCC Patients

We investigated the prognostic significance of the deep learning based TASIL-score for DSS of HNSCC patients. First, we considered C1 as a discovery cohort and both C2 and C3 as a joint validation cohort. Patients in the validation cohort were divided into two groups based on the TASIL-score using an optimal threshold value obtained from the analysis of the discovery cohort. We found that the patient group with the higher TASIL-score shows significantly better DSS (*p*=0.000003, hazard ratio [HR] = 0.20, 95% confidence interval [CI] 0.10–0.43) on the validation cohort. The TASIL-score was again statistically significant in patient stratification for DSS (*p*=0.00239, HR = 0.49, 95% CI 0.30–0.78) when the discovery and validation cohorts were swapped. The Kaplan-Meier curves along with the corresponding log-rank test based *p*-values are presented in Figure 2. These curves show a clear separation between low- and high-risk patient groups when stratified using the TASIL-score.

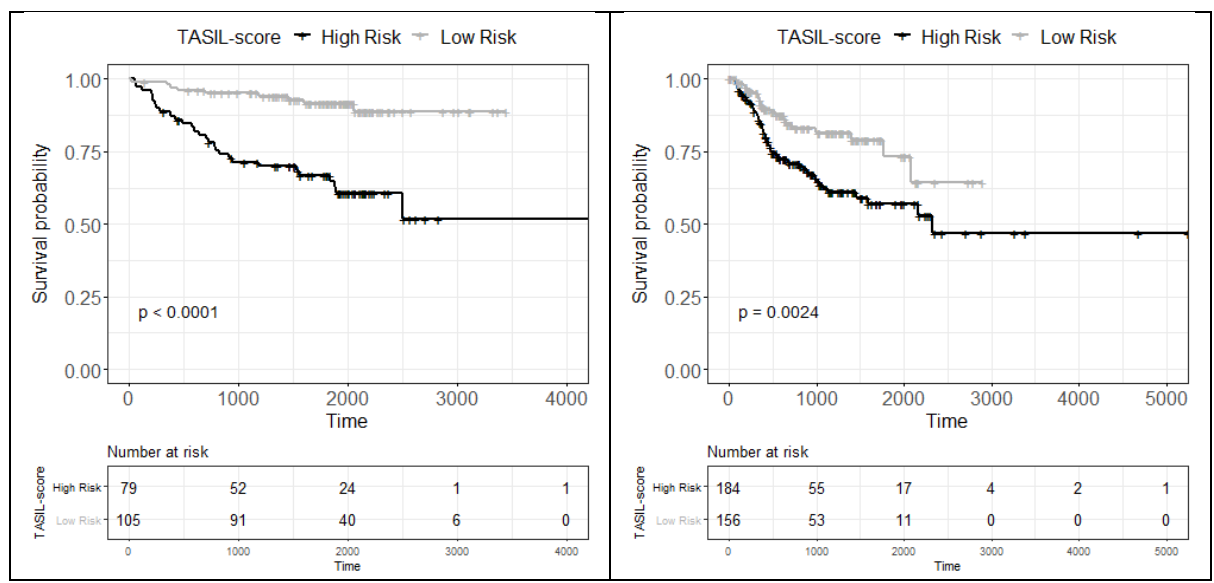

*Figure 2: Kaplan-Meier curves of low- and high-risk patients in C2 & C3 (left) and C1 (right) cohorts when used as validation cohorts for disease-specific survival.*

## TASIL-score is a prognostic indicator for the SCC of both oral and oropharynx sites

The C2 and C3 cohorts were curated from only the oral cavity and oropharynx, respectively. Therefore, we investigated the prognostic significance of TASIL-score for patients with SCC of a specific site. First, the OSCC cohort (C2) was considered as a discovery cohort and the OPSCC cohort (C3) was considered as a validation cohort. Supporting our aforementioned findings, the TASIL-score remains prognostically significant (*p*=0.000159, HR = 0.20, 95% CI 0.08–0.49) for OPSCC patient stratification into low- and high-risk groups for DSS. Second, we swapped the two cohorts considering the OPSCC cohort (C3) as the discovery cohort and the OSCC cohort (C2) as a validation cohort. We found that the TASIL-score based OSCC patient stratification again proved prognostically significant (*p*=0.000935, HR = 0.08, 95% CI 0.01–0.65). Repeating the experiments to evaluate the prognostic significance of the TASIL-score for DFS follows the same pattern where the TASIL-score stratifies patients in

TASIL-score predicts survival in head and neck squamous cell carcinoma prognostically significant low- and high-risk groups. Patient stratification into low- and high-risk groups is presented in Figure 3 through Kaplan-Meier curves for all four experiments.

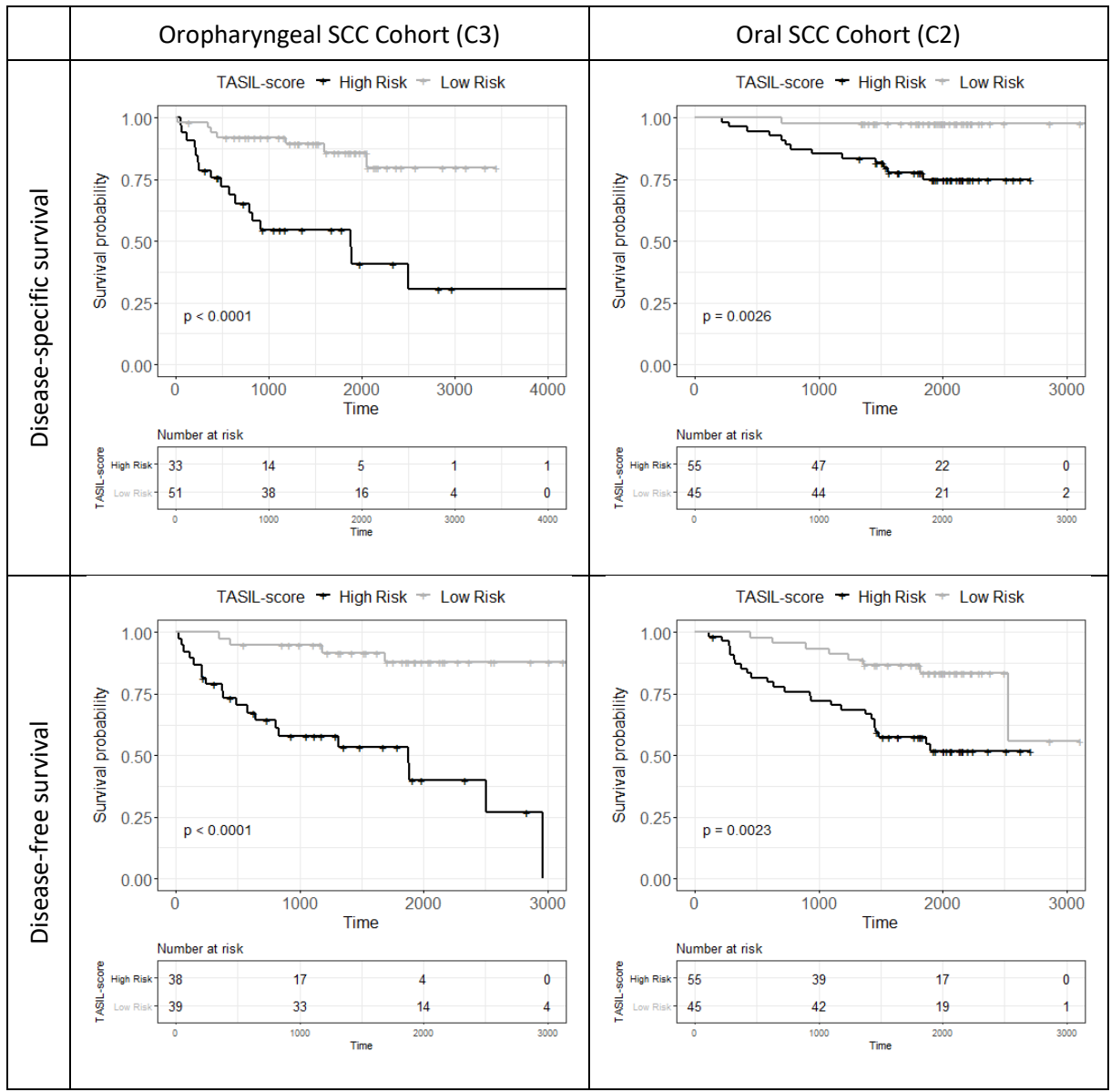

*Figure 3: Kaplan-Meier curves of low- and high-risk patients in Oropharyngeal (C3) and Oral (C2) cohorts when used as validation cohorts for disease-specific and disease-free survival.*

## TASIL-score is independent of clinical and pathological variables

We investigated the prognostic significance of the TASIL-score through multivariate analysis in the presence of clinical and pathological variables. The C2 cohort was considered for multivariate analysis as it has more clinical and pathological parameters along with both DSS and DFS information (Table S2) compared to the other two cohorts. For DSS, both TASIL-score ($p$=0.027, HR = 0.10, 95% CI 0.01–0.76) and the pathological stage ($p$=0.043, HR = 2.02, 95% CI 1.02–3.97) were found to be independent prognostic variables against all other variables (Table 1). However, for DFS, TASIL-score is the only independent variable of statistical significance ($p$=0.004, HR = 0.29, 95% CI 0.12–0.67) against age, sex, smoke and smokeless tobacco status, tumour grade, patterns of invasion, and pathological stage. We also investigated the prognostic significance of the TASIL-score for DSS in a multivariate setting on the C1 cohort using the available clinical and pathological variables. The TASIL-score remains prognostic ($p$=0.043, HR = 0.58, 95% CI 0.34–0.98) in the presence of other clinicopathological variables: age, gender, grade, and pathological stage. Although stage IVb and IVc appear to be statistically significant, the total number of patients in stage IVb and IVC are 9 and 1, respectively, which is quite small compared to the total number of patients (Figure 4).

*Table 1: Multivariate analysis of TASIL-score in the presence of available clinical and pathological variables of SKM cohort for both disease-specific and disease-free survival.*

| Variables | Disease-specific Survival | | | | Disease-free Survival | | | |
|---|---|---|---|---|---|---|---|---|
| | HR | 95% CI | | $p$-value | HR | 95% CI | | $p$-value |
| | | Lower | Upper | | | Lower | Upper | |
| Age | 1.015 | 0.969 | 1.060 | 0.526 | 1.010 | 0.980 | 1.040 | 0.579 |
| Sex | 2.057 | 0.502 | 8.430 | 0.316 | 1.540 | 0.650 | 3.660 | 0.323 |
| Smoking Tobacco | 1.905 | 0.505 | 7.190 | 0.342 | 1.160 | 0.510 | 2.690 | 0.72 |
| Smokeless Tobacco | 0.671 | 0.156 | 2.900 | 0.593 | 1.990 | 0.770 | 5.130 | 0.155 |
| Tumour Grade | 0.689 | 0.312 | 1.520 | 0.355 | 1.090 | 0.660 | 1.790 | 0.73 |
| Invasion Pattern | 1.121 | 0.653 | 1.920 | 0.678 | 1.350 | 0.930 | 1.950 | 0.115 |
| TNM Stage | 2.015 | 1.021 | 3.970 | **0.043** | 1.090 | 0.780 | 1.520 | 0.627 |
| TASIL | 0.099 | 0.013 | 0.760 | **0.027** | 0.290 | 0.120 | 0.670 | **0.004** |

TASIL-score predicts survival in head and neck squamous cell carcinoma

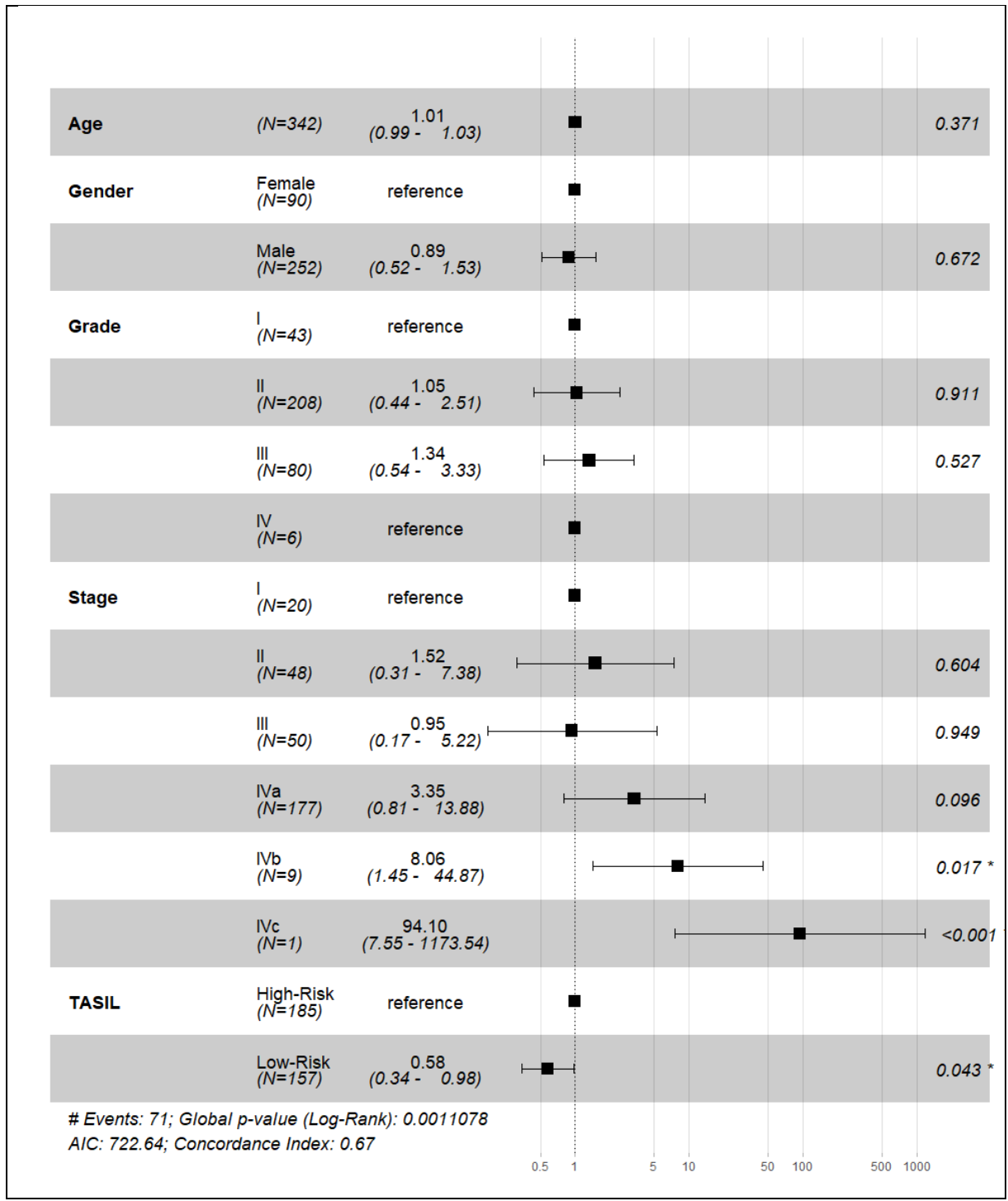

*Figure 4: Multivariate analysis of TASIL-score in the presence of available clinical and pathological variables of C1 cohort.*

## TASIL-score shows a better separation between low- and high-risk patients compared to the manual TIL score

The pathologist score for tumour/stroma infiltrating lymphocytes is usually a categorical score with low, moderate, and high infiltration categories. Only low and high categories show prognostic significance in the C3 cohort for DSS and DFS, as shown in Table S3 . We merge the moderate category in two different ways to split the patients into low- and high-risk groups, see Figure 5. The Kaplan-Meier curves in Figure 5 show that the TASIL-score stratifies patients into low- and high-risk groups with better prognostic significance compared to the manual TIL score in both DSS and DFS analyses.

## Comparison with existing digital scores

In recent years, researchers have proposed several automated quantification methods for colocalisation in different types of cancers to develop a digital prognostic biomarker [19], [21], [27], [28]. The use of computerised methods addresses the issue of subjectivity and produces objective and reproducible quantification scores. Geessink *et al.* [28] presented tumour to stroma ratio (TS-Ratio) for rectal adenocarcinoma and showed that a higher TS-Ratio had prognostic association with poor patient survival. Maley *et al.* [21] used an ecological measure for quantification of immune-cancer cell colocalisation (IC-Colocalisation) in breast cancer. They found that the higher colocalisation of immune and tumour cells was associated with better patient survival.

Similarly, Shaban *et al.* [19] proposed a tumour infiltrating lymphocytes abundance (TILAb) score for OSCC which showed prognostic significance for DFS in both univariate and multivariate analysis. We use Harrell's concordance index (C-Index) to compare the predictive ability of the TASIL-score with the existing automated quantification methods in six different experiments. In Figure 6, each set of bars represents the results on a cohort when used as validation. The proposed TASIL-score achieves the best C-Index scores in four experiments and comparable C-Index scores in the remaining two experiments for DSS and DFS.

TASIL-score predicts survival in head and neck squamous cell carcinoma

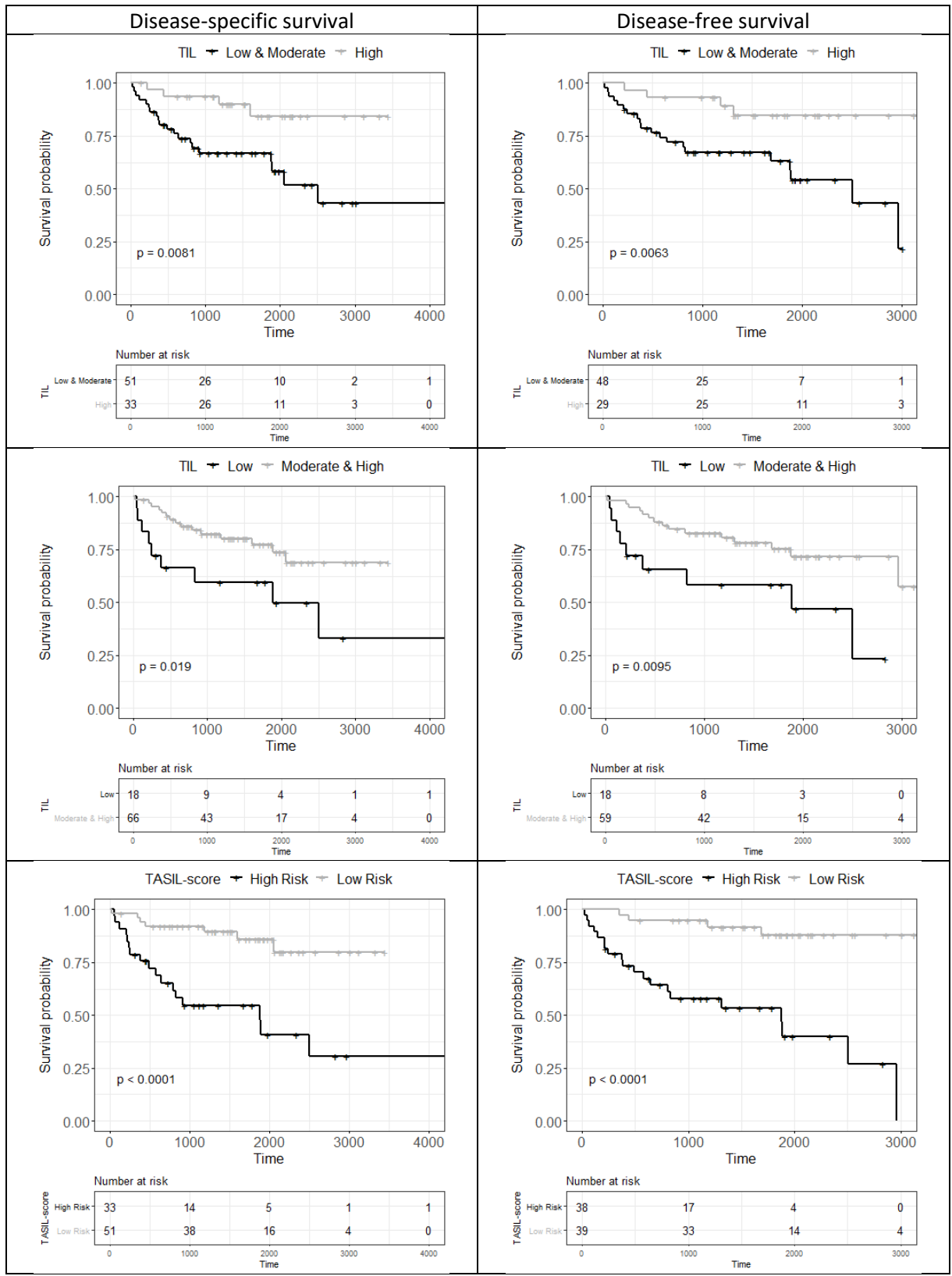

*Figure 5: Comparison of manual pathologist TIL score (1st row – Low & Moderate vs High and 2nd row for Low vs Moderate & High) and proposed TASIL-score (last row) in univariate setting for disease-specific and disease-free survival on C3 cohort.*

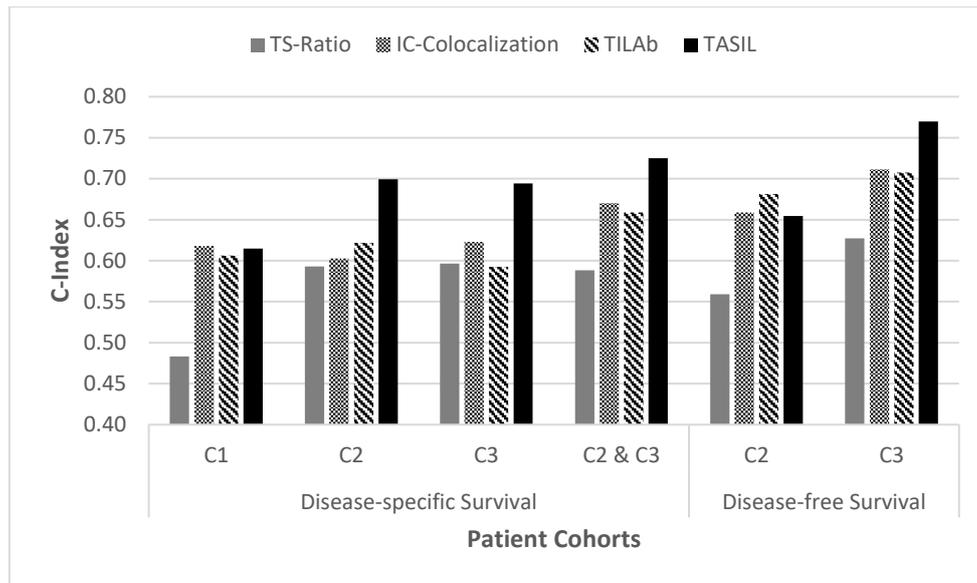

*Figure 6: Concordance index of different tumour, tumour-associated stroma and lymphocytes quantification methods, including the proposed TASIL-score.*

## TASIL-score is correlated with molecular estimates of CD8+ T cells

We further investigated the correlation of the proposed TASIL-score with molecular estimates of immune cell fractions in the TCGA-HN cohort (C1). Throsson *et al*. [29] have estimated the fraction of 22 immune cell types in the histology slides of patients in the TCGA cohort using gene expression data through CIBERSORT. We used those estimates for the correlation analysis with our TASIL-score. The immune subtypes were grouped based on nine different immune cell types: dendritic, mast, neutrophils, eosinophils, monocytes, macrophages, natural killer cells, T cells and B cells. Our main finding is that the TASIL-score shows a moderate but highly significant positive correlation with T cell estimates and negative correlation with macrophage estimates (Figure 7).

Moreover, CD8+ T cell fraction shows the highest positive correlation among all immune subtypes (Table 2), which may indicate that the lymphocytes in the vicinity of the TAS are mainly CD8+ T cells [30]. An explanation for the value of correlation between TASIL-score and cell fractions not being very high is that TASIL-score and the molecular estimates are computed on formalin-fixed paraffin-embedded (FFPE) and fresh frozen tissue sections, respectively. Both tissue sections belong to tissue blocks from the same patient, but their exact spatial relation is unknown. However, a good correlation of TASIL-score with CD8+ T cell indirectly validates the significance of the proposed score.

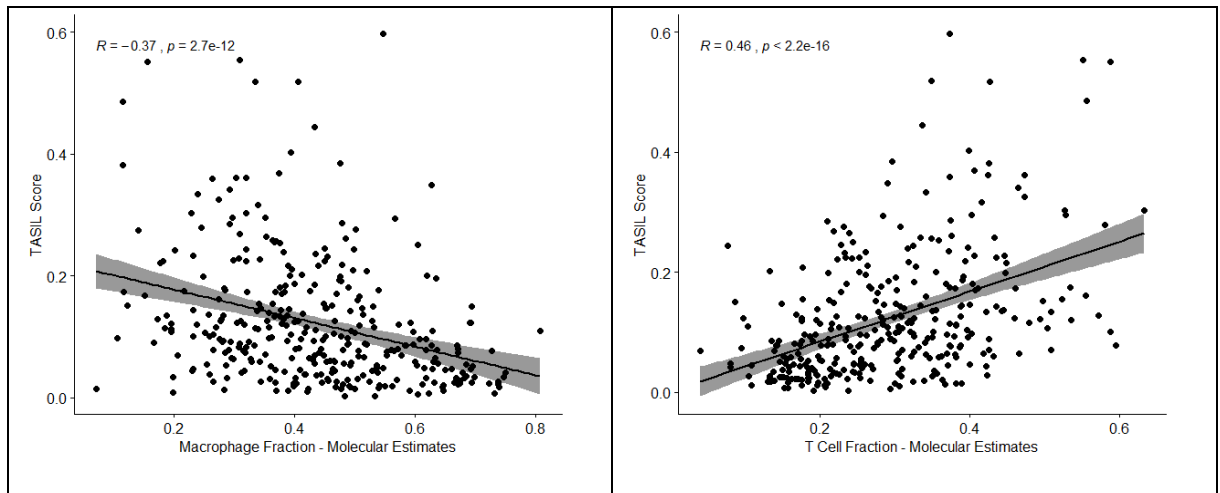

*Figure 7: Spearman correlation between TASIL-score and molecular estimates of Macrophages (left) and T Cells (right) fractions.*

*Table 2: Spearman correlation between TASIL-score and molecular estimates of immune subtypes.*

| Cell Types | rho | *p*-value | Cell Subtypes | rho | *p*-value |
|---|---|---|---|---|---|
| Mast Cells | -0.18 | $8.22 \times 10^{-4}$ | Activated | -0.18 | $1.19 \times 10^{-3}$ |
|  |  |  | Resting | 0.12 | $2.84 \times 10^{-2}$ |
| Monocytes | 0.18 | $7.70 \times 10^{-4}$ | Monocytes | 0.18 | $7.70 \times 10^{-4}$ |
| Macrophages | -0.37 | $2.67 \times 10^{-12}$ | M0 | -0.39 | $2.08 \times 10^{-13}$ |
|  |  |  | M1 | 0.32 | $2.62 \times 10^{-9}$ |
|  |  |  | M2 | -0.24 | $1.05 \times 10^{-5}$ |
| T-Cells | 0.46 | $1.32 \times 10^{-18}$ | CD4 Memory Activated | 0.32 | $2.04 \times 10^{-9}$ |
|  |  |  | CD4 Memory Resting | -0.09 | $9.66 \times 10^{-2}$ |
|  |  |  | CD4 Naive | -0.20 | $3.11 \times 10^{-4}$ |
|  |  |  | CD8 | 0.45 | $9.28 \times 10^{-18}$ |
|  |  |  | Follicular Helper | 0.26 | $2.37 \times 10^{-6}$ |
|  |  |  | Gammadelta | 0.07 | $1.94 \times 10^{-1}$ |
|  |  |  | Regulatory | 0.26 | $1.16 \times 10^{-6}$ |
| B-Cells | 0.22 | $3.97 \times 10^{-5}$ | Memory | -0.02 | $7.37 \times 10^{-1}$ |
|  |  |  | Naive | 0.22 | $4.32 \times 10^{-5}$ |
|  |  |  | Plasma | 0.11 | $4.77 \times 10^{-2}$ |

## Discussion

We have proposed a deep learning based objective measure, namely the TASIL-score, for quantification of tumour-associated stroma infiltrating lymphocytes in digitised images of HNSCC tissue slides. We found that higher value of the TASIL-score was associated with better DSS of HNSCC patients (Figure 2) and with both DSS and DFS of OSCC and OPSCC patients (Figure 3). The TASIL-score was independent of clinicopathological parameters for DSS and DFS of OSCC patients (Table 1). It also showed better separation between low- and high-risk OPSCC patients compared to manual TIL score assessed by an expert pathologist (Figure 5). We compared the TASIL-score with the existing automated quantification methods for stroma and lymphocytic quantification with respect to the tumour. The TASIL-score achieves high concordance score compared to its counterparts (Figure 6) and also showed a moderate but highly significant correlation with molecular estimates of CD8+ T cells (Table 2).

Most of the existing automated quantification methods were developed for the quantification of lymphocytes or stroma in relation to the tumour such as stroma to tumour ratio in breast and ovarian cancer [27], [28], lymphocyte and tumour colocalisation in breast cancer [21], and abundance of tumour infiltrating lymphocyte in OSCC [19]. However, automated quantification of stromal TILs has not been fully explored yet. To the best of our knowledge, the proposed TASIL-score is the first automated quantitative score of lymphocytic infiltration in the TAS of HNSCC.

The role of TASIL has been investigated in several clinical studies for different cancers [10], [23], [24]. Salgado *et al.* [10] found that stromal TILs are a superior and more reproducible prognostic parameter compared to intratumoral TILs in breast cancer. Xu *et al.* [24] also reported that stromal TILs were of clinical relevance as a high stromal TIL score was associated with better patient prognosis for HNSCC. Furthermore, stromal TILs have been shown to be an independent risk factor for DSS and DFS of HNSCC patients. Our automated TASIL-score has shown similar prognostic significance pattern, see Figure 2 and Figure 3.

In clinical practice, both H&E and immunohistochemistry (IHC) staining are used for manual TIL scoring. The TASIL-score based results are in agreement with previous findings based on H&E and IHC based TIL quantification [23], [31], [32]. In Figure 5, we have shown that both manual H&E based pathologist score and TASIL-score carry prognostic significance for DSS and DFS of OPSCC. However, TASIL-score shows better separation between Kaplan-Meier curves of low- and high-risk patients of OPSCC. Ruiter *et al.* [31] conducted a meta-analysis of IHC based studies to investigate the prognostic value of T cells in HNSCC. They found a favourable prognostic role of CD3+ and CD8+ T cell infiltration in HNSCC patients. Balermpas *et al.* [23] found that high CD8 expression in tumour stroma is a prognosticator for HNSCC. The proposed TASIL-score also shows a positive and highly significant correlation with genomic estimates of CD8+ T cells in HNSCC patients in the TCGA-HNSCC cohort.

T lymphocyte infiltration in the stroma and tumour indicates an effective immune challenge to the tumour and is related to better outcome and treatment response. The proposed TASIL-score based findings are aligned with the clinical knowledge with the added advantages of objectivity, reproducibility, and strong prognostic value. Therefore, the automated, objective, and quantitative TASIL-score has the potential to provide valuable insights into tumour behaviour and prognosis in an efficient and consistent manner. Although we validated our method on three different cohorts (n=537 cases in total), a comprehensive evaluation on large multicentric cohorts is required before the proposed digital score can be adopted in clinical practice.

# Supplementary Materials

Patient Characteristics

In TCGA-HN cohort (C1), most of the patients were diagnosed between 2007 to 2013, and the average age of the patients is 61.09 year with a standard deviation of 11.82. There are more male patients (n=252) compared to female patients (n=90). The distribution of TNM-stage of the cases is somewhat skewed toward higher stages, with 52% cases from stage IVa patients. The ratio of alive and deceased patients is also imbalanced, with only 25% deceased cases, and the average DSS of all patients is 28.69 months with a standard deviation of 24 months. Detailed statistics of the cohort are presented in Table S1.

In the SKM cohort (C2), all the patients were diagnosed between 2010 to 2013, and the average age of the patients is approximately 50 years, with 11.12 years of standard deviation. The grade, growth pattern and the pathologists' manual TIL score information along with the TNM stage is available for almost all the patients. The most dominated stage, grade and growth pattern are IVA, II and II, respectively. The low and moderate TIL groups show prognostic significance for both overall and diseases free survival. In terms of survival, most patients were alive until the last follow-up; however, 32 patients suffered from disease recurrence. Table S2 presents a detailed description of all the available parameters of the cohort and their prognostic significance, if applicable.

Patients in the PredicTR1 cohort (C3) were diagnosed between 2000 to 2010 and were tracked until 2014. The cases in this cohort were collected from six different centres with a minimum of nine and a maximum of 23 cases from a centre. The DSS data is available for 84 cases, whereas disease-free information is available for 77 cases. There are 24 recurrent and 25 deceased cases out of those with available survival data. TILs are manually scored into only three categories: low, moderate, and high. Presence of lymphocytes in 80% or more of tumour/stroma is categorised as high TILs and the presence of lymphocytes in less than 20% of tumour/stroma is denoted by low TILs. Table S3 presents a detailed description of all the available parameters of the cohort along with the prognostic significance where applicable.

## Pathologist Annotations

We used 24 cases, one WSI per case, for training and evaluation of the coarse segmentation method. Half of the cases are taken from the C1 cohort, and the remaining half were selected from C2 cohort. Multiple visual fields of size 256×256 at 10× magnification (280×280 µm) were extracted from each case for multi-class tissue annotation by the expert pathologist. The pathologist then assigned a label to each 32×32 (35×35 µm) regions, 64 per image, in all the images from the predefined set of seven classes: Tumour, Lymphocyte/Inflammatory, Stroma, Keratin, Epithelium, Artifacts, and Other for remaining tissue regions. All the annotated visual fields were then split into training and validation sets where all visual fields from a case lied only in one set. Training and validation sets consisted of 141,541 and 38,893 annotated regions, respectively. Table S4 presents a detailed distribution of the annotated regions in each set.

## Whole Slide Image Segmentation

The architecture our patch based segmentation is presented in Figure S3 which consists of multiple convolution, pooling, and feature concatenation layers. All the convolution layers preceded by a batch norm and ReLU based activation layers apart from the first one where these two layers are used after the convolution layer. The main building block of the proposed network is the dense block which consists of multiple pair of convolution layers where the depth of a block depends on the number of iterations selected for the block. In each iteration, the pair of convolution layers converts the input feature-map into 32 channels feature-map and concatenate it with the input feature-map. The last convolution layer followed by softmax layer takes the output feature-maps of all the dense blocks through skip connections after spatial average pooling, if required, and outputs the probability maps for the given number of classes.

We compared our segmentation method with existing patch-based segmentation methods using average accuracy and macro F1-score based metrics. Our proposed segmentation method achieved a superior performance compared to standard patch-based segmentation methods (see Table S5). Visual results of our segmentation method are presented in Figure S4, which shows good segmentation of some of the main constituents of the tumour microenvironment.

# Supplementary Tables

*Table S1: Summary of available parameters of the C1 cohort along with log-rank test based p-values for disease-specific survival.*

| Categorical Parameters | | Count | Percentage | *p*-value |
|---|---|---|---|---|
| Number of Cases | | 342 | 100% | - |
| Gender | Male | 252 | 74% | 0.401 |
| | Female | 90 | 26% | |
| TNM Stage | I | 20 | 6% | 0.143 |
| | II | 48 | 14% | 0.085 |
| | III | 50 | 14% | **0.021** |
| | IVa | 177 | 52% | **0.003** |
| | IVb&c | 10 | 3% | **0.003** |
| | Not Reported | 37 | 11% | - |
| Disease-specific Status | Alive | 255 | 74.6% | - |
| | Deceased | 85 | 24.8% | |
| | Not Reported | 2 | 0.6% | |
| Continuous Parameters | | Mean | Standard Dev | *p*-value |
| Age (years) | | 61.09 | 11.8 | 0.647 |
| Disease-specific Survival (Months) | | 28.69 | 24.0 | - |

TASIL-score predicts survival in head and neck squamous cell carcinoma

*Table S2: Summary of available parameters of the C2 cohort along with log-rank test based p-values for disease-specific survival (DSS) and disease-free survival (DFS).*

| Categorical Parameters | | Count | Percentage | DSS *p*-value | DFS *p*-value |
|---|---|---|---|---|---|
| Number of Cases | | 100 | 100% | - | - |
| Sex | Male | 57 | 57% | 0.003 | 0.196 |
| | Female | 43 | 43% | | |
| Smoking Tobacco | Yes | 32 | 32% | 0.138 | 0.191 |
| | No | 66 | 66% | | |
| | Not Reported | 2 | 2% | | |
| Smokeless Tobacco | Yes | 19 | 19% | 0.823 | 0.197 |
| | No | 79 | 79% | | |
| | Not Reported | 2 | 2% | | |
| Tumour Grade | I | 35 | 35% | 0.484 | 0.540 |
| | II | 50 | 50% | 0.990 | 0.769 |
| | III | 15 | 15% | 0.363 | 0.688 |
| Invasion Pattern | I | 18 | 18% | 0.250 | 0.079 |
| | II | 18 | 18% | 0.212 | 0.156 |
| | III | 35 | 35% | 0.238 | 0.240 |
| | IVa | 28 | 28% | 0.209 | 0.081 |
| | Not Reported | 1 | 1% | - | - |
| TNM Stage | I | 25 | 25% | **0.020** | 0.183 |
| | II | 14 | 14% | 0.982 | 0.738 |
| | III | 15 | 15% | 0.813 | 0.914 |
| | IVa | 43 | 43% | **0.021** | 0.124 |
| | Not Reported | 3 | 3% | - | - |
| Disease-specific Status | Alive | 86 | 86% | - | |
| | Deceased | 14 | 14% | | |
| Disease Recurrence | Yes | 32 | 32% | - | |
| | No | 86 | 68% | | |
| Continuous Parameters | | Mean | Standard Dev | DSS *p*-value | DFS *p*-value |
| Age (years) | | 49.57 | 11.1 | 0.861 | 0.364 |
| Survival (Months) | Disease-specific | 60.1 | 17.8 | - | |
| | Disease-free | 53.56 | 22.3 | | |

Table S3: Summary of available parameters of the C3 cohort along with log-rank test based p-values for disease-specific survival (DSS) and disease-free survival (DFS).

| Categorical Parameters | | Count | Percentage | DSS *p*-value | DFS *p*-value |
|---|---|---|---|---|---|
| Number of Cases | | 95 | 100% | - | - |
| Sex | Male | 61 | 64% | 0.130 | 0.239 |
| | Female | 34 | 36% | | |
| Pathologist's TIL Score | Low | 22 | 23% | **0.019** | **0.009** |
| | Moderate | 37 | 39% | 0.414 | 0.440 |
| | High | 36 | 38% | **0.008** | **0.006** |
| Disease-specific Status | Alive | 59 | 62% | - | |
| | Deceased | 25 | 26% | | |
| | Not Reported | 11 | 12% | | |
| Disease Recurrence | Yes | 24 | 25% | - | |
| | Moderate | 53 | 56% | | |
| | No | 18 | 19% | | |
| Continuous Parameters | | Mean | Standard Dev | DSS *p*-value | DFS *p*-value |
| Age (years) | | 57.74 | 11.5 | 0.065 | 0.134 |
| Survival (Months) | Disease-specific | 47.22 | 30.3 | - | |
| | Disease-free | 47.25 | 29.1 | | |

Table S4: Distribution of annotated regions (sub-images or patches) for each class in both training and validation sets.

| Classes | Training | Validation | Total |
|---|---|---|---|
| Tumour | 35,627 | 12,863 | 48,490 |
| Lymphocyte | 10,488 | 5736 | 16,224 |
| Tumour-associated Stroma | 15,248 | 2161 | 17,409 |
| Keratin | 6735 | 3552 | 10,287 |
| Epithelium | 17,884 | 4354 | 22,238 |
| Others | 39,331 | 7450 | 46,781 |
| Artefacts | 16,228 | 2777 | 19,005 |
| Total | 141,541 | 38,893 | 180,434 |

Table S5: Comparison of different patch-based WSI segmentation methods.

| Methods | Accuracy (%) | F1_Score |
|---|---|---|
| ResNet-50 | 76.15 | 0.7323 |
| MobileNet-1.0 | 76.62 | 0.7478 |
| DenseNet-121 | 81.25 | 0.7876 |
| **Ours** | **85.11** | **0.8311** |

# Supplementary Figures

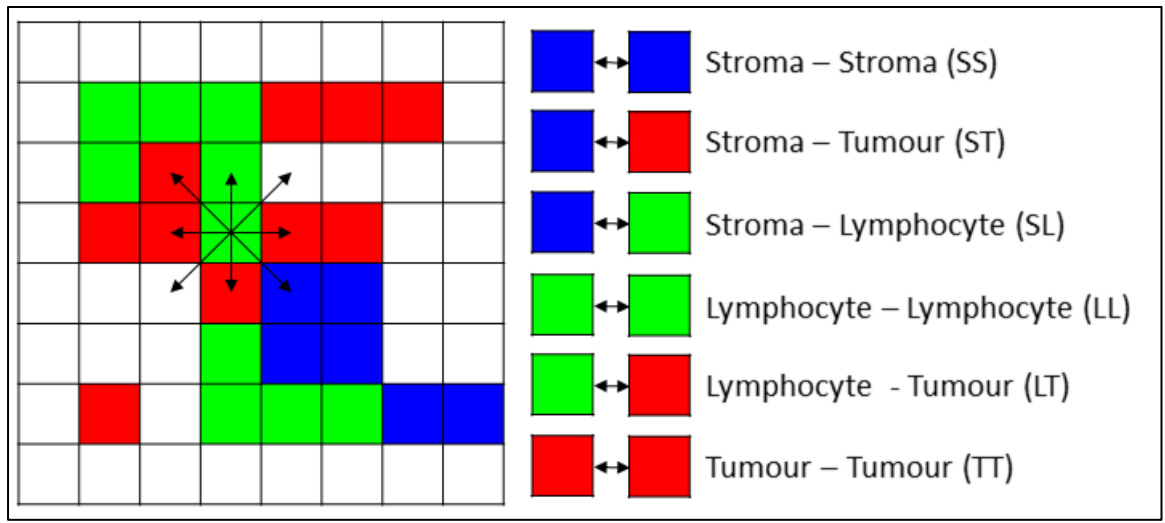

*Figure S 1: Visual illustration of patch co-occurrence analysis: Right half of the figure lists the six different patch co-occurrence patterns that can appear in the image (left). Here, stroma refers to tumour-associated stroma for the sake of brevity.*

TASIL-score predicts survival in head and neck squamous cell carcinoma

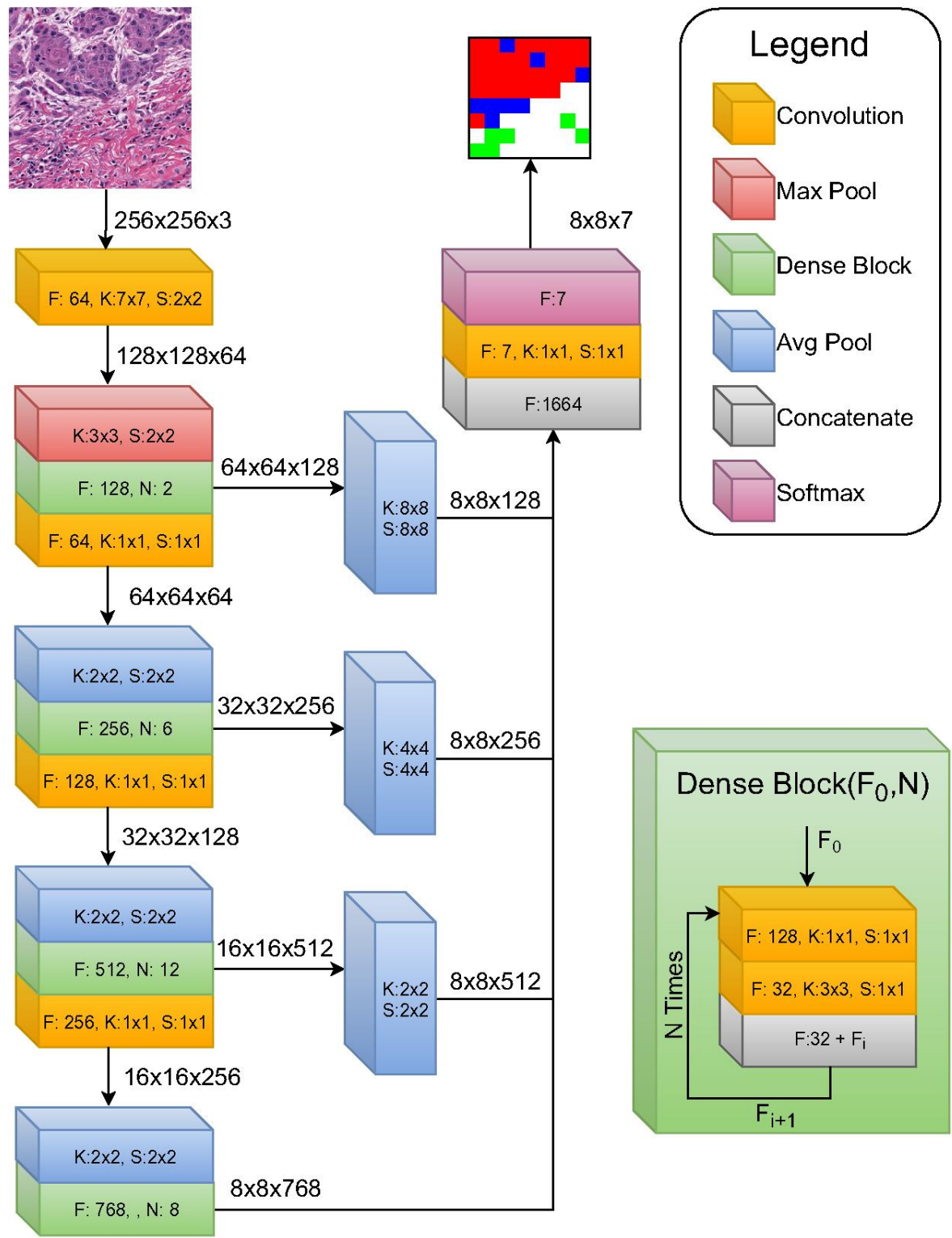

*Figure S 2: The architecture of the segmentation network using DenseNet as a baseline. Each box in the prediction map represents the prediction of a 32×32 corresponding region in the input patch. The letters N, F, K, and S represent the dense block depth, output feature maps, kernel size, and stride size, respectively.*

TASIL-score predicts survival in head and neck squamous cell carcinoma

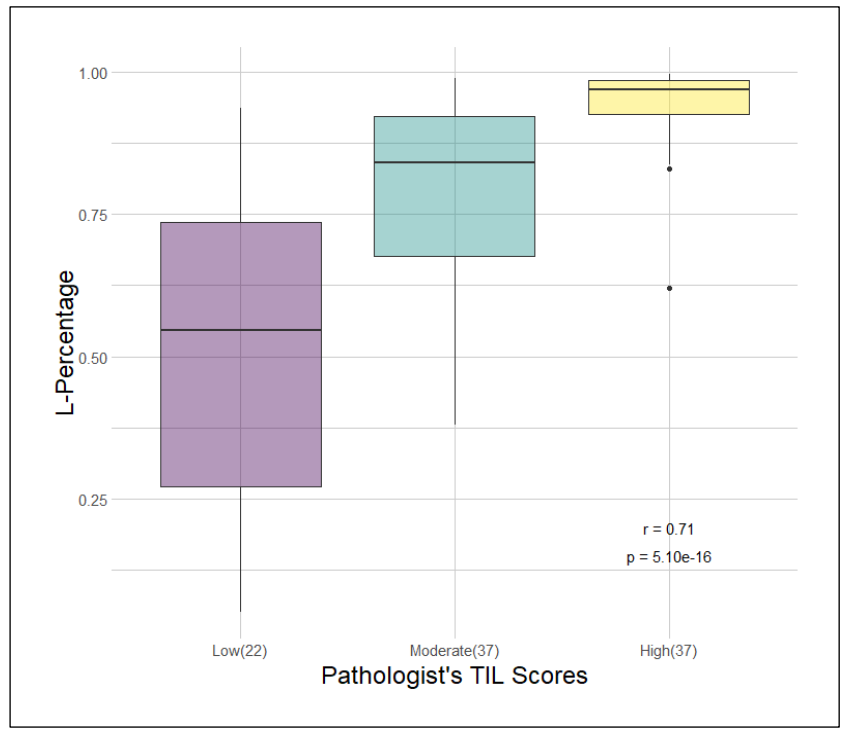

*Figure S 3: Boxplot representation of L-Percentage and pathologists' manual TIL score for all cases in the C3 cohorts. Boxplot representation of L-Percentage and pathologists' manual TIL score for all cases in the C3 cohorts. The number of cases in each TIL score group are presented in parentheses. Spearman corelation (r) and its statistical significance (p-value) are presented at bottom right of the figure.*

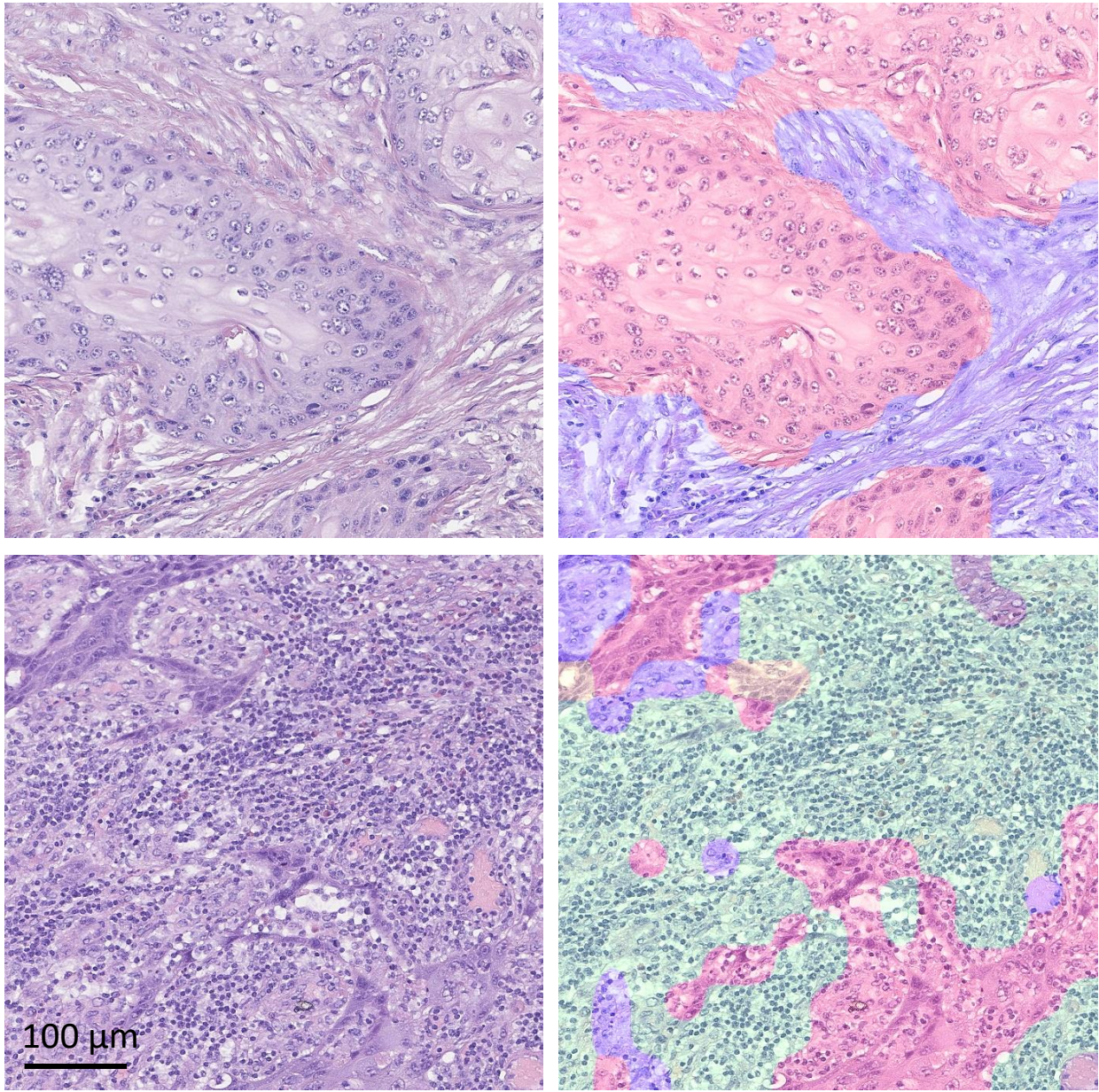

*Figure S 4: Visual results of our segmentation method. The right column shows the overlay of predicted class in different colours where tumour, tumour-associated stroma, lymphocyte, and Non-ROI regions are represented by red, blue, green, and yellow colours, respectively.*